\documentstyle[12pt,epsfig,axodraw,a4]{article} 
\def\bild#1#2{    
        \vspace*{-5mm}
        \begin{center}
        \begin{math}
        \epsfxsize#2cm
        \epsffile{#1}
        \end{math}
        \end{center}
        }
\newcommand{\vs}{\vspace{-0.25cm}}

\begin{document} 

\begin{center}
\large{\bf Chiral 3\begin{boldmath}$\pi$\end{boldmath}-exchange NN-potentials:
Results for diagrams proportional to $g_A^4$ and $g_A^6$}

\medskip 

\bigskip

N. Kaiser\\

\bigskip

Physik Department T39, Technische Universit\"{a}t M\"{u}nchen,\\
    D-85747 Garching, Germany

\end{center}

\bigskip

\begin{abstract}
We calculate in (two-loop) chiral perturbation theory the local NN-potentials
generated by the three-pion exchange diagrams proportional to $g_A^4$ and
$g_A^6$. Surprisingly, we find that the total isoscalar central 
$3\pi$-exchange potential vanishes identically. The individually largest 
$3\pi$-exchange potentials are of isoscalar spin-spin, isovector central and
isoscalar tensor type. For these potentials simple analytical
expressions can be given. The strength of these dominant $3\pi$-exchange
potentials at $r=1.0\,$fm is 4.6\, MeV, 2.9\,MeV and 1.4\,MeV, respectively.
Furthermore, we observe that the spin-spin and tensor potentials due to the
diagrams proportional to $g_A^6$ do not exist in the infinite nucleon mass 
limit. 
\end{abstract}


\bigskip
\bigskip

In a recent work \cite{3pipot} we have started to calculate the static 
nucleon-nucleon potentials generated by the (two-loop) $3\pi$-exchange 
diagrams with interaction vertices taken from the leading-order effective 
chiral $\pi N$-Lagrangian. 
In ref.\cite{3pipot} we restricted ourselves to the evaluation of four 
relatively simple classes of diagrams which in fact comprise all graphs with 
the common prefactor $g_A^2/f_\pi^6$. It was stressed in ref.\cite{3pipot} 
that the chiral $3\pi NN$-contact vertex depends on the choice of the 
interpolating pion-field and therefore one has to consider representation 
invariant classes of diagrams (classes I and II in ref.\cite{3pipot}) by 
supplementing graphs involving the chiral $4\pi$-vertex. Another obvious 
consequence of this is that there is no unique way (in chiral perturbation 
theory) of defining the so-called "pion-nucleon form factor" which is often 
introduced in phenomenological models. The only meaningful separation of the
interaction potential is the one into point-like $1\pi$-exchange and 
$3\pi$-exchange. The four classes of chiral $3\pi$-exchange diagrams evaluated
in ref.\cite{3pipot} gave rise only to isovector spin-spin  and tensor
potentials. The strength of the resulting coordinate-space potentials was very
weak with values of  $\pm 0.1$\,MeV  and less at a
internucleon distance of $r=1.0\,$fm. Compared to the chiral $2\pi$-exchange
potentials \cite{nnpap1,nnpap2} these are negligibly small corrections. 

The purpose of this work is complete the calculation of the chiral 
$3\pi$-exchange NN-potentials and to present results for the diagrams carrying
independent prefactors $g_A^4/f_\pi^6$ and $g_A^6/f_\pi^6$. We will see that a
grouping of these diagrams into five different classes is very useful.

Let us begin with some basic definitions in order to fix our notation. In the
static limit and considering only irreducible diagrams the on-shell NN T-matrix
takes the following form: 
\begin{eqnarray} {\cal T}_{NN} &=& V_C(q) + V_S(q)\,\vec \sigma_1\cdot \vec 
\sigma_2 + V_T(q) \,\vec \sigma_1\cdot \vec q \,\,  \vec \sigma_2 \cdot \vec q
\nonumber \\ && + \Big[ W_C(q) + W_S(q)\,\vec \sigma_1\cdot  \vec \sigma_2
+ W_T(q) \,\vec \sigma_1\cdot \vec q \,\,  \vec \sigma_2 \cdot \vec q \, \Big] 
\,\vec \tau_1 \cdot \vec \tau_2\,\,, \end{eqnarray} 
where $q=|\vec q\,|$ denotes the momentum transfer between the initial and
final state nucleon. The subscripts $C,S$ and $T$ refer to the central, 
spin-spin and tensor components, each of which occurs in an isoscalar $(V)$ and
an isovector $(W)$ version. As indicated, the (real) NN-amplitudes $V_C(q),
\dots, W_T(q)$ depend only on the momentum transfer $q$ in the static
limit. We are here interested only in the coordinate-space potentials generated
by certain diagrams in which three pions are simultaneously exchanged between 
both nucleons. For this purpose it is sufficient to calculate the imaginary 
parts of the NN-amplitudes $V_{C,S,T}(q)$ and $W_{C,S,T}(q)$ analytically 
continued to  time-like momentum transfer $q=i\mu-0^+$ with $\mu\geq 3m_\pi$. 
These imaginary parts are then the mass-spectra entering a representation of 
the local coordinate space potentials in form of a continuous superposition of
Yukawa  functions,     
\begin{equation} \widetilde V_C(r) = -{1\over 2\pi^2 r} \int_{3m_\pi}^\infty
d\mu \,\mu \,e^{-\mu r} \, {\rm Im}\, V_C(i\mu)\,\,, \end{equation}
\begin{equation} \widetilde V_S(r) = {1\over 6\pi^2 r} \int_{3m_\pi}^\infty
d\mu \,\mu \,e^{-\mu r} \Big[ \mu^2\, {\rm Im}\, V_T(i\mu) - 3\, {\rm Im}\, 
V_S(i\mu) \Big]\,\,, \end{equation} 
\begin{equation} \widetilde V_T(r) = {1\over 6\pi^2 r^3} \int_{3m_\pi}^\infty
d\mu\,\mu\, e^{-\mu r}(3+3\mu r+\mu^2r^2){\rm Im}\,V_T(i\mu)\,\,.\end{equation}
The isoscalar central, spin-spin and tensor potentials, denoted here by
$\widetilde V_{C,S,T}(r)$, are as usual those ones which are accompanied by the
operators $1$, $\vec \sigma_1\cdot \vec \sigma_2$ and  $3\,\vec \sigma_1\cdot
\hat r\,\vec \sigma_2 \cdot \hat r -\vec \sigma_1\cdot \vec \sigma_2$,
respectively. For the isovector potentials $\widetilde W_{C,S,T}(r)$ a
completely analogous representation holds. 

Let us now turn to the evaluation of the NN-potentials from the two-loop
$3\pi$-exchange diagrams. Application of the Cutkosky cutting rules gives the
relevant imaginary parts Im\,$V_{C,S,T}(i\mu)$ and Im\,$W_{C,S,T}(i\mu)$ as 
integrals of the squared $\bar NN \to 3\pi$ transition amplitudes over the 
Lorentz-invariant three-pion phase space. Some details about these techniques 
can be found in ref.\cite{bkm} where the same method has been applied to
calculate (two-loop) spectral-functions of nucleon form factors. The pertinent
three-body space integrals are most conveniently performed in the three-pion
center-of-mass frame. The corresponding on-mass-shell four-momenta of the three
pions read in this frame: $k_1^\nu = (\omega_1, \vec k_1\,)$, $k_2^\nu = 
(\omega_2, \vec k_2\,)$ and $k_3^\nu = (\mu-\omega_1-\omega_2, -\vec k_1-\vec
k_2)$. The mass-shell condition $k_3^2=m_\pi^2$ determines the cosine of the
angle between $\vec k_1$ and $\vec k_2$ (called $z$) as 
\begin{equation} z\, k_1k_2 = \omega_1\omega_2 -\mu(\omega_1+\omega_2)
+{1\over2} (\mu^2+m_\pi^2)\,\,, \qquad k_{1,2} = \sqrt{\omega^2_{1,2}-m_\pi^2}
\,\,. \end{equation} 
In the chosen reference frame the four-momenta of the external nucleons
can be taken as $P_1^\nu = (\mu/2, p\, \vec v\,)$ and $P_2^\nu = (-\mu/2, p\,
\vec v\,)$, where $\vec v$ is a (real) unit vector and $p=\sqrt{\mu^2/4-M^2} = 
i\, M +{\cal O}(M^{-1})$ in the heavy nucleon limit ($M\to \infty$). Of course 
the assignment of $P_1^\nu$ and $P_2^\nu$ to the external nucleon lines has to
be done such that four-momentum conservation holds for the transition 
$\bar N N \to 3\pi \to \bar N N$.  The  possible nucleon propagators occuring 
in diagrams are always of the form
\begin{equation} {2M \over (P_1-k_j)^2-M^2} = {1 \over i\, \vec v\cdot \vec k_j
- \epsilon} \,\,,\qquad {2M \over (P_2+k_j)^2-M^2} = {-1 \over i\, \vec v \cdot
  \vec k_j + \epsilon} \,\,. \end{equation}
The sign of the $\epsilon$-terms in eq.(6) is uniquely fixed from the original 
relativistic expressions and in the heavy nucleon mass limit $(M\to \infty)$
the positive quantity $\epsilon$ becomes infinitesimally small. With these 
techniques at hand it is possible to perform analytically at least two of the 
four non-trivial $3\pi$-phase-space integrations \cite{bkm}. As a check on
the heavy nucleon formalism employed we recalculated  the imaginary parts of
the isoscalar electromagnetic and isovector axial nucleon form factors and 
found perfect agreement with the results of ref.\cite{bkm}. As a further check
we applied the methods to calculate the imaginary parts of the $2\pi$-exchange
NN-amplitudes. In that case only a much simpler two-body phase space integral
needs to be evaluated and one reproduces indeed exactly the results of
ref.\cite{nnpap1,nnpap2} in a rather short calculation. 

\bigskip
\bigskip

\bild{3pipotfig5.epsi}{16}
\smallskip
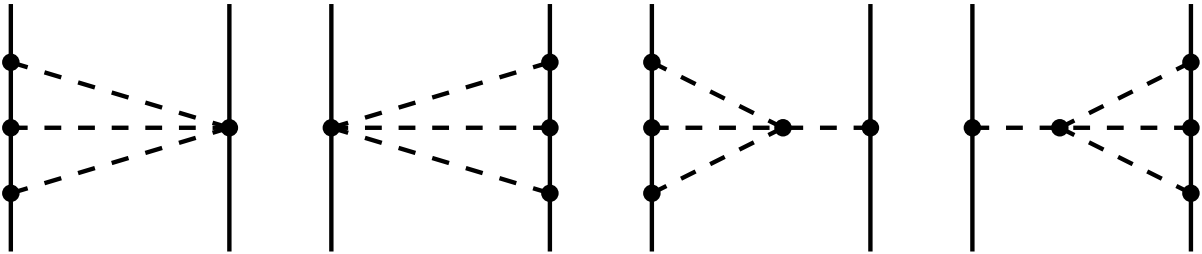
{\it Fig.1: $3\pi$-exchange diagrams of class\,V proportional to $g_A^4$. 
Solid and dashed lines represent nucleons and pions, respectively.}
\bigskip

Let us now turn to the results for the (two-loop) chiral $3\pi$-exchange
diagrams proportional to $g_A^4$ and $g_A^6$. We start with the diagrams of
class\,V shown in Fig.\,1. As stressed in ref.\cite{3pipot} diagrams involving
the chiral $3\pi NN$-vertex or the chiral $4\pi$-vertex depend on an arbitrary
parameter  $\alpha$ and therefore one should consider the full class\,V as one
entity. Obviously, the last two pion-pole diagrams in Fig.\,1 contribute via
coupling constant renormalization also to the point-like $1\pi$-exchange. This
effect is however automatically taken care by working with the physical $\pi
NN$-coupling constant $g_{\pi N}$. From an inspection of the spin- and isospin
factors occurring in the diagrams of class\,V one finds immediately that only 
non-vanishing isovector spin-spin and tensor NN-amplitudes $W_{S,T}$ will be 
obtained. We find the following imaginary parts from class\,V,     
\begin{equation} {\rm Im}\,W_S^{(V)}(i\mu) = {2g_A^4 \over 3(8\pi f_\pi^2)^3}
\int\!\!\!\! \int\limits_{\!\!\!\!z^2\leq1} \!\!\!d\omega_1d\omega_2 \bigg\{ 
k_1^2+\mu \omega_1 +3 (m_\pi^2 -\mu \omega_1) \Big( z+{k_2 \over
k_1} \Big) {\arccos(-z) \over \sqrt{1-z^2} } \bigg\} \,\,, \end{equation}
\begin{eqnarray} {\rm Im}\,W_T^{(V)}(i\mu) &=& {1\over \mu^2 } {\rm Im}\,
W_S^{(V)}(i\mu) +{g_A^4(\mu^2-m_\pi^2)^{-1} \over \mu^2(8\pi f_\pi^2)^3}
\int\!\!\!\! \int\limits_{\!\!\!\!z^2\leq1} \!\!\!d\omega_1d\omega_2 \nonumber
\\ && \times \Big[(6\mu^2+2m_\pi^2) (\omega_1+\omega_2)-\mu (4\mu^2+3m_\pi^2)
\Big] \nonumber\\ & & \times \bigg\{\Big[ (\mu^2+m_\pi^2) \Big( 2\omega_1-{\mu
\over 2} \Big) -2\mu \omega_1 \omega_2 \Big] {\arccos(-z) \over k_1 k_2
\sqrt{1-z^2} } +\mu +2z \omega_1  {k_2 \over k_1} \bigg\} \,\,. \end{eqnarray}
The inequality $z^2\leq 1$ defines the kinematically allowed (Dalitz) region in
the $\omega_1\omega_2$-plane (which is bounded by a cubic curve) together with
the  obvious kinematical
constraints $m_\pi \leq \omega_{1,2}\leq \mu-2m_\pi$ and $2m_\pi \leq \omega_1+
\omega_2 \leq \mu-m_\pi$. Note that the same integrand as in eq.(7) for
Im\,$W_S^{(V)}(i\mu)$ was found  in ref.\cite{bkm} for the spectral-function 
of the nucleon isovector axial form factor. In the chiral limit $m_\pi = 0$ one
can evaluate the remaining double-integrals in eqs.(7,8) using the substitution
$\omega_1 = \mu(1-xy)/2$, $\omega_2 = \mu y/2$ which maps the unit-square 
$0\leq x,y\leq 1$ onto the (in the chiral limit) triangle-shaped Dalitz
region.

One finds from class\,V repulsive isovector spin-spin and tensor
potentials with an $r^{-7}$-dependence,    
\begin{equation} \widetilde W_S^{(V)}(r)\Big|_{m_\pi=0} = { g_A^4 \, r^{-7} 
\over (4\pi)^5 f_\pi^6} \bigg( {16\over21} \pi^2+{85\over 36}\bigg)\,\,, 
\qquad \widetilde W_T^{(V)}(r)\Big|_{m_\pi=0} = { g_A^4 \, r^{-7} \over(4\pi)^5
 f_\pi^6} \bigg({245\over 72}\bigg) \,\,. \end{equation}

Next, we come to the diagrams of class\,VI shown in Fig.\,2. The isospin factor
of the first and second graph is $4 \vec \tau_1\cdot \vec \tau_2-6$ while
that of the third and fourth graphs is $-4 \vec \tau_1\cdot \vec\tau_2-6$.
The last two graphs are irreducible whereas the first two contain the 
iteration of $1\pi$-exchange and  $2\pi$-exchange (triangle subgraphs). This
iterative part has to be separated off in the construction of the NN-potential.
A closer inspection reveals that the two types of diagram differ only by some
signs in nucleon propagators. The first two diagrams carry a factor
$-[(i\vec v \cdot \vec k_1+\epsilon)(i\vec v \cdot \vec k_1-\epsilon)]^{-1}$ in
comparison to a factor $(i\vec v \cdot \vec k_1+\epsilon)^{-2}$ from the last
two (irreducible) diagrams. The irreducible part of the first two diagrams is
obtained by switching the sign of one $\epsilon$-term ($-\epsilon \to
+\epsilon)$ such that the expression agrees with that of the irreducible
diagrams. 
\bigskip
\bigskip

\bild{3pipotfig6.epsi}{16}
\smallskip
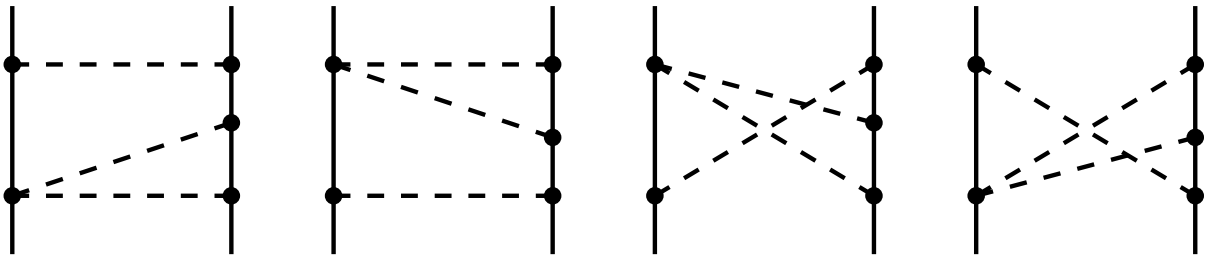
{\it Fig.2: $3\pi$-exchange diagrams of class\,VI proportional to $g_A^4$. 
Diagrams for which the role of both nucleons is interchanged are not shown. 
They lead to the same contribution to the NN-potential. The (irreducible) 
isoscalar NN-amplitudes sum up to zero.}
\bigskip
 
In order to make this procedure more understandable consider the
following integrals: $\int_{-1}^1 dx [(x+i\epsilon)(x-i\epsilon)]^{-1} = \pi/
\epsilon-2 +{\cal O}(\epsilon^2)$ and $\int_{-1}^1 dx(x+i\epsilon)^{-2} = -2
+{\cal O}(\epsilon^2)$. The difference between both diverges as $1/\epsilon$
for $\epsilon \to 0^+$. According to the definition in eq.(6) one has 
$\epsilon \sim M^{-1}$ and from the (non-relativistic) Lippmann-Schwinger 
equation \cite{erwe} it is known that the iteration of the potential leads to a
contribution proportional to the nucleon mass $M$. Therefore the 
$1/\epsilon$-term which gets subtracted by switching the sign (of one 
$\epsilon$) corresponds indeed to the iterative part. In the case of
$2\pi$-exchange all this has been worked out in detail in
ref.\cite{nnpap1} and as already mentioned the present methods allow to 
reproduce exactly the results of ref.\cite{nnpap1} for the irreducible 
$2\pi$-exchange. 

After subtracting the iterative part the first two and the last two diagrams 
in Fig.\,2 become equal up to a minus-sign. Combining this with the isospin 
factors one obtains again only a non-vanishing contribution to the isovector 
spin-spin and tensor NN-amplitudes $W_{S,T}$. We find the following imaginary
parts from class\,VI,  
\begin{equation} {\rm Im}\,W_S^{(VI)}(i\mu) = {2g_A^4 \over (8\pi f_\pi^2)^3}
\int\!\!\!\! \int\limits_{\!\!\!\!z^2\leq1} \!\!\!d\omega_1d\omega_2 \bigg\{ 
-k_1^2-{5\over 3} \mu \omega_1 + (\mu \omega_1-m_\pi^2 ) \Big( z+{k_2 \over
k_1} \Big) {\arccos(-z) \over \sqrt{1-z^2} } \bigg\} \,\,, \end{equation}
\begin{eqnarray} && {\rm Im}\,W_T^{(VI)}(i\mu) = {1\over \mu^2 } {\rm Im}\,
W_S^{(VI)}(i\mu) + {2g_A^4 \over \mu^2(8\pi f_\pi^2)^3} \int\!\!\!\!
\int\limits_{\!\!\!\!z^2\leq1} \!\!\!d\omega_1d\omega_2 \, \omega_1 \bigg\{
{2\omega_1 \over 3k_1^2}(2\mu \omega_1-\mu^2 +3m_\pi^2 -6\omega_2^2) \nonumber
\\ &&\qquad +\Big[(\mu^2+m_\pi^2) (\mu-2\omega_1- 2\omega_2) +4\mu \omega_1
\omega_2 \Big] {\arccos(-z) \over k_1 k_2 \sqrt{1-z^2} } -2z\omega_2{k_2\over 
k_1} +3\omega_1 -2\mu \bigg\} \,\,. \end{eqnarray}  
In the chiral limit ($m_\pi = 0)$ one gets now a repulsive isovector spin-spin
potential and an attractive isovector tensor potential of the form,
 
\begin{equation} \widetilde W_S^{(VI)}(r)\Big|_{m_\pi=0} = { g_A^4 \, r^{-7} 
\over (4\pi)^5 f_\pi^6} \bigg({175\over36} \bigg)\,\,, \qquad
\widetilde W_T^{(VI)}(r)\Big|_{m_\pi=0} ={ g_A^4 \, r^{-7} \over (4\pi)^5 
f_\pi^6} \bigg({4\over 3} \pi^2-{665\over 36}\bigg)\,\,. \end{equation}

Next, we come the diagrams of class\,VII shown Fig.\,3. The isospin factor of
all four diagrams is $6$ and after considering their spin-structure one finds
immediately that there will be only a non-vanishing contribution to the
isoscalar spin-spin and tensor NN-amplitudes $V_{S,T}$. In the case of 
class\,VII one can actually solve all integrals analytically and one obtains 
the following closed form  expressions for the imaginary parts,  
\bigskip
\bigskip

\bild{3pipotfig7.epsi}{16}
\smallskip
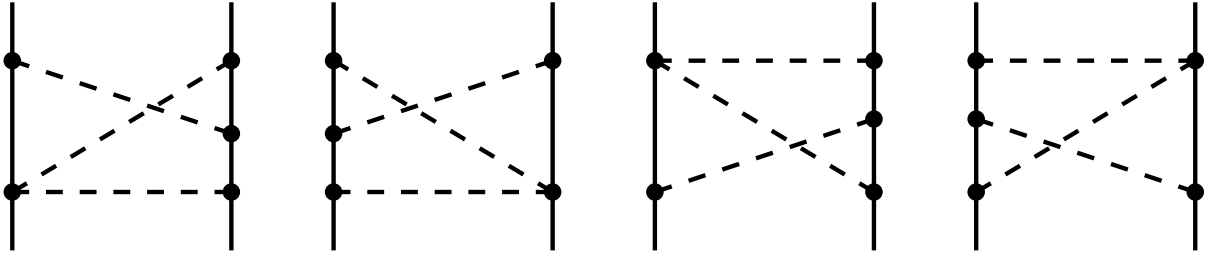
{\it Fig.3: $3\pi$-exchange diagrams of class\,VII proportional to $g_A^4$. 
The isospin factor of these diagrams is 6.}

\begin{equation} {\rm Im}\,V^{(VII)}_S(i\mu) = {g_A^4 (\mu-3m_\pi)^2 \over 35
\pi(32f_\pi^3)^2 } \bigg[ 2m_\pi^2 -12\mu m_\pi-2\mu^2+15 {m_\pi^3 \over \mu}+2
{m_\pi^4 \over \mu^2}+3{m_\pi^5 \over \mu^3} \bigg] \,\,, \end{equation}
\begin{equation} {\rm Im}\,V_T^{(VII)}(i\mu) = {g_A^4 (\mu-3m_\pi) \over 35\pi
(32\mu f_\pi^3)^2 } \bigg[ \mu^3 +3\mu^2m_\pi +2\mu m_\pi^2 +6m_\pi^3+18
{m_\pi^4 \over \mu}-9{m_\pi^5 \over \mu^2}-27{m_\pi^6 \over \mu^3} \bigg] \,. 
\end{equation}
It is even more astonishing that the corresponding coordinate space potentials
(inserting eqs.(13,14) into eqs.(3,4)) can be expressed through a simple
exponential-function multiplied by a polynomial. We find the following
repulsive isoscalar spin-spin and tensor potentials from class\,VII,    
\begin{equation} \widetilde V_S^{(VII)}(r) = { g_A^4 \over 2(8\pi f_\pi^2)^3}
{e^{-3m_\pi r} \over r^7} (1+m_\pi r)^2(2+m_\pi r)^2\,\,,\end{equation}
\begin{equation} \widetilde V_T^{(VII)}(r) = {g_A^4 \over 2(8\pi f_\pi^2)^3}
{e^{-3m_\pi r} \over r^7}(1+m_\pi r)^2(1+m_\pi r+m_\pi^2r^2)\,\,.\end{equation}

\bigskip

\bild{3pipotfig8.epsi}{8}
\smallskip
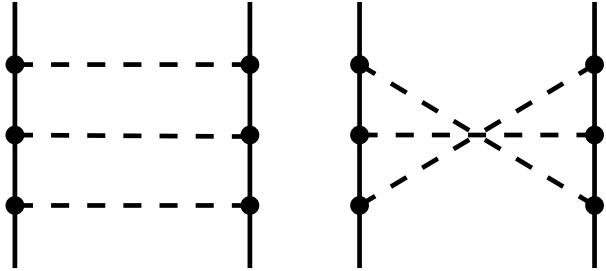
{\it Fig.4: $3\pi$-exchange diagrams of class\,VIII proportional to $g_A^6$.}

\bigskip

Next, we consider the diagrams of class\,VIII shown in Fig.\,4. The isospin
factor of the first graph is   $7 \vec \tau_1\cdot \vec \tau_2-6$ while that of
the second one is $7 \vec \tau_1\cdot \vec \tau_2+6$. In order to separate off
the iterative parts from the first diagram one has to switch the sign of two
different $\epsilon$-terms in nucleon propagators. After this procedure the
factors coming from the nucleon propagators agree identically for both diagrams
in Fig.\,4. The two diagrams differ however in the ordering of the first and 
third (spin-dependent) pion-coupling to one nucleon line. Exploiting
furthermore the properties of $\vec \sigma$-matrices one finds that the
spin-independent part of this particular $3\pi$-exchange interaction is equal
with opposite sign for both diagrams, whereas the spin-dependent part is equal
with the same sign. Combining this result with the isospin factors one obtains
only non-vanishing isoscalar central ($V_C$) and isovector spin-spin and tensor
($W_{S,T}$) NN-amplitudes. Explicit evaluation of the diagrams of class\,VIII 
leads to the following imaginary part of the isoscalar central NN-amplitude, 
\begin{equation} {\rm Im}\,V_C^{(VIII)}(i\mu) = -{3g_A^6 \mu^2 \over 2(8\pi 
f_\pi^2)^3} \int\!\!\!\! \int\limits_{\!\!\!\!z^2\leq1}\!\!\!d\omega_1d\omega_2
\Big\{ 1+ {z \over \sqrt{1-z^2} } \arccos(-z) \Big\} \,\,. \end{equation}
In the chiral limit ($m_\pi=0)$ one finds a repulsive isoscalar central
potential with an $r^{-7}$-dependence of the following form, 
\begin{equation} \widetilde V_C^{(VIII)}(r)\Big|_{m_\pi=0} = { g_A^6 \, r^{-7}
\over (4\pi)^5 f_\pi^6} \Big( 60-4\pi^2\Big)\,\,.\end{equation}
A similar calculation gives for the imaginary part of the isovector spin-spin 
NN-amplitude $W_S$ generated by  class\,VIII the following result,    
\begin{eqnarray} {\rm Im}\,W_S^{(VIII)}(i\mu) &\stackrel{?}{=}& {14g_A^6 \over
(16\pi f_\pi^2)^3} \int\!\!\!\! \int\limits_{\!\!\!\!z^2\leq1} \!\!\!d\omega_1d
\omega_2 \bigg\{ {1\over 3}(\omega_1^2-\mu^2-4m_\pi^2 +8\mu \omega_1) \nonumber
\\ & &+\bigg[z+{\arccos(-z)\over\sqrt{1-z^2}}\bigg] {\mu\omega_1 -m_\pi^2\over 
k_1 (1-z^2) } \bigg[ {z\over k_1}(m_\pi^2-\mu \omega_1) +{m_\pi^2-\mu 
\omega_2\over k_2} \bigg] \bigg\} \,\,. \end{eqnarray}
However, as indicated by the question mark this results is of no use. On the
boundary of the Dalitz region, $z^2=1$, the integrand in eq.(19) has 
singularities of the form $(1-z^2)^{-1}$ and $(1-z^2)^{-3/2}$ such that the
double-integral $\int\!\!\int_{z^2\leq1} d\omega_1d\omega_2$ diverges. 
The same non-integrable singularities were found in an explicit calculation of
Im\,$W_T^{(VIII)}(i\mu)$. Therefore one has to conclude that the isovector 
spin-spin and tensor potentials generated by the diagrams of class\,VIII do not
exist in the infinite nucleon mass ($M\to \infty)$ or static limit. Note that
this statement applies separately to the second (irreducible) diagram in
Fig.\,4 and the occurrence of these singularities has nothing to do with adding
the (irreducible components of the) first diagram. Presently, we do not have
a detailed understanding of the problem encountered here. Most likely it 
originates from the fact that several nucleon propagators in the infinite mass 
limit $M\to \infty $ can introduce severe singularities in loop-functions as 
the following one-loop example shows,
\begin{equation} \int {d^4 l\over i(2\pi)^4 } { 1\over (l_0-\omega)^3
(m_\pi^2-l^2) } = {1\over 8\pi^2} \bigg[ {\omega\over m_\pi^2-\omega^2 } 
+{m_\pi^2 \over (m_\pi^2-\omega^2)^{3/2}} \arccos{-\omega \over m_\pi}
\bigg]\,. \end{equation}
The discovery of the non-finiteness of the isovector spin-spin and tensor 
potentials due to class\,VIII in the static ($M\to \infty$) limit points 
towards a further limitation of the heavy baryon approach when applied to the
two-nucleon system. It seems that once the number of heavy nucleon propagators
exceeds a critical value the corresponding two-loop integrals do not exist
anymore (for $M\to \infty$). Of course a deeper understanding of this feature
would be very desirable.

Finally, we consider the diagrams of class\,IX shown in Fig.\,5. The isospin
factor of the first and second diagram is $-6 -\vec \tau_1\cdot \vec \tau_2$
while that of the third and fourth diagram is $6-\vec \tau_1\cdot \vec \tau_2$.
The iterative parts of the last two diagrams are separated off in the standard
way by switching one sign of an $\epsilon$-term in a single nucleon propagator.
After a lengthy calculation we find for the imaginary part of the isoscalar 
central NN-amplitude $V_C$ from class\,IX, 
\begin{equation} {\rm Im}\,V_C^{(IX)}(i\mu) = {3g_A^6 \mu^2 \over 2(8\pi 
f_\pi^2)^3} \int\!\!\!\! \int\limits_{\!\!\!\!z^2\leq1}\!\!\!d\omega_1d\omega_2
\bigg\{ 1+ {k_2 \over k_1 \sqrt{1-z^2} } \Big[{\pi \over 2}-2 \arccos(-z) \Big]
  \bigg\} \,\,. \end{equation} 

\bigskip

\bild{3pipotfig9.epsi}{16}
\smallskip
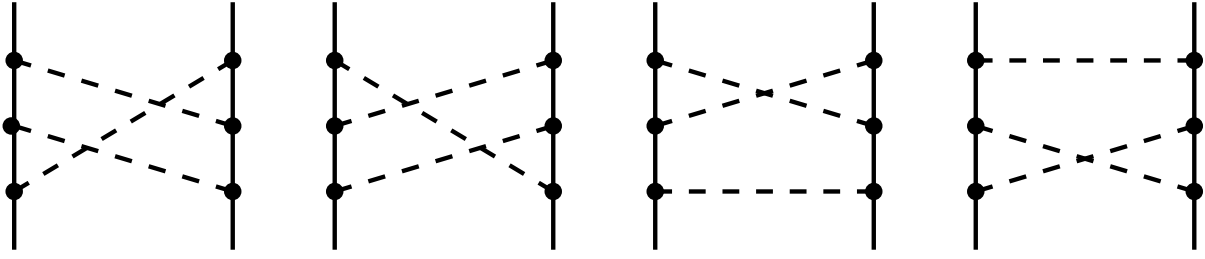
{\it Fig.5: $3\pi$-exchange diagrams of class\,IX proportional to $g_A^6$.}
\bigskip

In order to arrive at the very short expression given in eq.(21) we have 
exploited  the inherent permutational symmetry of the $3\pi$-phase space 
integration. In the chiral limit $m_\pi=0$ one obtains from class\,IX an 
attractive  isoscalar central potential of the form
\begin{equation}\widetilde V_C^{(IX)}(r)\Big|_{m_\pi=0} ={ g_A^6 \, r^{-7} 
\over (4\pi)^5 f_\pi^6} \Big(4\pi^2-60\Big)\,\,, \end{equation}
which just cancels the one coming from class\,VIII given in eq.(18). Later when
presenting numerical results in Table\,1 we will see that this cancellation 
holds also for any finite pion mass $m_\pi$. Furthermore, we obtain from 
class\,IX a non-vanishing contribution to the isovector central NN-amplitude
$W_C$. For the corresponding imaginary part all appearing integrals can be
solved analytically and one finds the following closed form expression,
\begin{equation} {\rm Im}\,W^{(IX)}_C(i\mu) = {g_A^6 (\mu-3m_\pi)^2 \over 30
\pi \mu(4f_\pi)^6 } ( 3m_\pi^3 +2\mu m_\pi^2-9\mu^2m_\pi-4\mu^3) \,\,.
\end{equation}
Inserting this mass-spectrum into eq.(2) one obtains finally a repulsive 
isovector central potential which can be expressed in terms of a simple
exponential-function multiplied by a polynomial,   
\begin{equation} \widetilde W_C^{(IX)}(r) = { 2g_A^6 \over (16\pi f_\pi^2)^3}
{e^{-3m_\pi r}\over r^7}(1+m_\pi r)(4+5m_\pi r+3m_\pi^2r^2)\,\,.\end{equation} 
We have also evaluated the imaginary parts of the isoscalar and isovector
spin-spin and tensor NN-amplitudes arising from the four diagrams shown in 
Fig\,5. In these cases there appear again singular terms proportional
to $(1-z^2)^{-3/2}$  in the integrands such that the double-integral 
$\int\!\!\int_{z^2\leq1} d\omega_1d\omega_2$ diverges. Consequently, we may 
conclude that only the central potentials generated by the $3\pi$-exchange
diagrams proportional to $g_A^6$ (shown in Figs.\,4,5) exist in the static
limit ($M\to \infty$) while all spin-spin and tensor potentials diverge. This
is a somewhat unexpected feature of the $3\pi$-exchange NN-potential. 
  
In Table\,1, we present numerical results for the coordinate space 
NN-potentials generated by the $3\pi$-exchange graphs of class\,V, VI, VII,
VIII and IX for internucleon distances 0.6\,fm$\,\leq r\leq \,$1.4\,fm. We use
the parameters $f_\pi=92.4$\,MeV, $m_\pi=138\,$MeV (average pion mass) and
$g_A=g_{\pi N}f_\pi/M=1.32$ employing the Goldberger-Treiman relation together
with $g_{\pi N}=13.4$. The choice $g_A=1.32$ is most natural in the present
context since the pion-nucleon coupling is the relevant here and not the 
axial-vector coupling. One observes that the isovector spin-spin and tensor
potentials from classes\,V and VI (scaling with $g_A^4$) are a factor 3 to 10
larger than the (largest) potential $\widetilde W_{S,T}^{(III)}(r)$ found in 
our previous work \cite{3pipot} on the chiral $3\pi$-exchange diagrams scaling
with $g_A^2$. The most striking result is that the isoscalar central potentials
from class\,VIII and IX cancel each other, $\widetilde V_C^{(VIII)}(r)+
\widetilde V_C^{(IX)}(r)=0$, and this cancellation is found to happen with high
numerical precision. From the double-integral representation in eqs.(17,21) it
is not at all obvious that the relation Im\,$V_C^{(VIII)}(i\mu)+{\rm Im}\,
V_C^{(IX)}(i\mu)=0$ holds, at least we have not yet found a simple analytical
proof. Consequently, we can conclude that the total isoscalar central potential
generated by chiral $3\pi$-exchange vanishes identically. Only the exchange of
an explicit $\omega(782)$-vector meson (a strongly resonant $3\pi$-state) can 
contribute to the (phenomenologically needed) isoscalar central repulsion.

By inspection of Table\,1 we furthermore observe that the largest chiral
$3\pi$-exchange potentials are the repulsive isoscalar spin-spin and tensor 
potentials $\widetilde V_{S,T}^{(VII)}(r)$ due to class\,VII and the repulsive
isovector central potential $\widetilde W_C^{(IX)}(r)$ due to class\,IX. These
are just the cases in which simple analytical expressions could be given 
in eqs.(15,16,24). Actually, we are disproving here the claim of 
ref.\cite{robil} that a particular diagram of class\,I (namely the first graph
in Fig.\,1 of ref.\cite{3pipot}) would be dominant.  Presumably, the strength
of the  potentials $\widetilde V_{S,T}^{(VII)}(r)$ and $\widetilde W_C^{(IX)}
(r)$ is still too weak to be of practical relevance. Of course, in order to 
confirm this an empirical analysis of the low-energy NN-data base as done in
ref.\cite{nijmeg} including these  dominant chiral $3\pi$-exchange potentials
would be very desirable.
    
In summary, we have completed in this work the calculation of the static chiral
$3\pi$-exchange NN-potentials by evaluating all diagrams proportional to 
$g_A^4$ and $g_A^6$. As a surprise we find that the total isoscalar central
potential vanishes identically. The individually largest chiral $3\pi$-exchange
NN-potentials are of isoscalar spin-spin, isovector central and isoscalar
tensor type and for these potentials very simple analytical expressions could
be  given. We furthermore discovered that the spin-spin and tensor potentials
due to class\,VIII and IX (consisting of diagrams proportional to $g_A^6$) do
not exist in the heavy nucleon mass limit $M\to \infty$.    

\begin{table}[hbt]
\begin{center}
\begin{tabular}{|c|ccccccccc|}
    \hline

    $r$~[fm]&0.6&0.7&0.8&0.9&1.0&1.1&1.2&1.3 &1.4\\
   \hline $\widetilde W_S^{(V)}$~[MeV]
&51.7&16.5& 6.02 &2.43&1.06&0.503& 0.249&0.129&0.069\\ 
 $\widetilde W_T^{(V)}$~[MeV]
& 16.0& 4.99& 1.79 &0.704&0.302&0.141&0.068&0.035& 0.018 \\ \hline    
$\widetilde W_S^{(VI)}$~[MeV]
&19.9 &5.99&2.06&0.783&0.322&0.141&0.065&0.031&0.015\\ 
    
$\widetilde W_T^{(VI)}$~[MeV]&--30.2&--9.78&--3.63 &--1.52&--0.689&--0.326&
--0.165&--0.087& --0.048\\ \hline
$\widetilde V_S^{(VII)}$~[MeV]&214.23 & 68.81& 25.38 & 10.40&4.62&2.19&1.10&
0.575&0.313\\
    
$\widetilde V_T^{(VII)}$~[MeV]& 58.39& 19.20& 7.25 &3.05&1.39&0.675&0.347&
0.186 &0.104\\    \hline
$\widetilde V_C^{(VIII)}$~[MeV]&177.6 &56.2& 20.4 &8.21&3.58&1.68& 0.830
&0.429 &0.229\\   \hline
$\widetilde V_C^{(IX)}$~[MeV]&--177.6 &--56.2 & --20.4 &--8.21&--3.58&--1.68
& --0.830 &--0.429 &--0.229\\    
$\widetilde W_C^{(IX)}$~[MeV] & 148.80& 46.54& 16.75& 6.70& 2.91& 1.36& 0.665&
0.342& 0.183 \\ 
   \hline
  \end{tabular}
\end{center}
{\it Tab.1: Numerical values of the local NN-potentials generated by the
chiral $3\pi$-exchange graphs of classes\,V,\,VI,\,VII,\,VIII,\,IX (shown in
Figs.\,1,\,2,\,3,\,4,\,5) versus the nucleon distance $r$. The
units of these potentials are MeV.}
\end{table}


\vskip-0.9 cm
\begin{thebibliography}{99} 
\bibitem{3pipot} N. Kaiser, {\it Phys. Rev.} {\bf C61} 0140xx (2000) in print;
nucl-th/9910044.\vs 
\bibitem{nnpap1} N. Kaiser, R. Brockmann and W. Weise, {\it Nucl. Phys. } {\bf
A625}, 758 (1997) and refs. therein.\vs
\bibitem{nnpap2} N. Kaiser, S. Gerstend\"orfer and W. Weise, {\it Nucl. Phys.} 
{\bf A637}, 395 (1998).\vs       
\bibitem{bkm} V. Bernard, N. Kaiser and U.G. Mei{\ss}ner, {\it Nucl. Phys.}
{\bf A611}, 429 (1996).\vs
\bibitem{erwe} T. Ericson and W. Weise, {\it Pions in Nuclei} 
(Clarendon Press, Oxford, 1988), app.\,10.\vs
\bibitem{robil} J.C. Pupin and M.R. Robilotta, {\it Phys. Rev.} {\bf C60},
014003 (1999).\vs    
\bibitem{nijmeg} M.C.M. Rentmeester, R.G.E. Timmermans, J.L. Friar and J.J. de
Swart, {\it Phys. Rev. Lett.} {\bf 82}, 4992 (1999).\vs 
\end{thebibliography}
\end{document}